\documentclass[prb,showpacs,preprint]{revtex4}   
\usepackage{xspace}
\usepackage{times,mathptm}                            
\usepackage{graphicx}                            
\usepackage{epstopdf}
\usepackage{amsmath}
\usepackage{amssymb}
\newcommand{\celsius}{$^{\circ}$C\xspace}

\begin{document}

\title{Quantum coherence of electrons in random networks of c-axis oriented wedge-shaped GaN nanowalls grown by molecular beam epitaxy}
\author{H.\ P.\ Bhasker$^{1}$}
\author{Varun\ Thakur$^{2}$, S.\ M.\ Shivaprasad$^{2}$}
\author{S.\ Dhar$^{1}$}
\email{dhar@phy.iitb.ac.in}
\affiliation{Physics Department, Indian Institute of Technology Bombay Powai, Mumbai 400076, India}
\affiliation{International Centre for Material Science, Jawaharlal Nehru Centre for Advanced Scientific Research, Bangalore 560064, India}

\begin{abstract}
The depth distribution of the transport properties as well as the temperature dependence of the low field magneto-conductance for several c-axis oriented GaN nanowall network samples grown with different average wall-widths ($t_{av}$) are investigated. Magneto-conductance recorded at low temperatures shows clear signature of weak localization effect in all nanowall samples studied here. The scattering mean free path $l_e$ and the phase coherence time $\tau_{\phi}$, are extracted from the magneto-conductance profile. Electron mobility estimated from $l_e$ is found to be comparable with those estimated previously from room temperature conductivity data for these samples\cite{bhasker1,bhasker2}, confirming independently the substantial mobility enhancement in these nanowalls as compared to bulk. Our study furthermore reveals that the high electron mobility region extends down to several hundreds of nanometer below the tip of the walls. Like mobility, phase coherence length ($l_{\phi}$) is found to increase with the reduction of the average wall width.  Interestingly, for samples with lower values of the average wall width, $l_{\phi}$ is estimated to be as high as 60 $\mu$m, which is much larger than those reported for GaN/AlGaN heterostructure based two dimensional electron gas (2DEG) systems.
\end{abstract}

\maketitle
Recently, electron mobility in networks of c-axis oriented wedge-shaped GaN nanowalls is estimated to be several orders of magnitude larger than that is observed in GaN bulk.\cite{bhasker1} Interestingly, the mobility has been found to increase with the reduction of the average width of the walls.\cite{bhasker2} The origin of these effects are not very clear yet. High mobility was speculated to be resulting from the transport of electrons through the edge states located at the top edges of the walls.\cite{bhasker1} It is, thus,  important to find out, whether the effect is confined only at the tip of the walls or not. Furthermore, keeping in mind that the system exhibits very high electron mobility(estimated to be as high as $\approx 10^{4}$~cm$^{2}$/$V-\sec$),  it will be interesting to study the quantum interference effect in this system. One of the important parameters of quantum interference is the phase coherence length  ($l_{\phi}$), that is the average length scale over which the phase information of an electron remains intact. In fact, there are many device proposals utilizing quantum interference effect.\cite{Datta} However, the device size has to be less than $l_{\phi}$, which is one of the key factors for the realization of these devices. Weak localization effect (WL) is often utilized to estimate $l_{\phi}$.\cite{bergmann,Kramer} WL is a negative correction to the conductance arises due to the enhancement of the effective scattering cross-section as a result of constructive quantum interference between a closed path of electron and its time reversed path. When a magnetic field is applied, it reduces the negative correction by removing the phase coherence between the two paths \cite{bergmann}, resulting in a positive magneto-conductance at low fields.   

Here, we have investigated the quantum coherence effect  in several c-axis oriented GaN nanowall network samples grown with different average wall-widths ($t_{av}$) by studying the temperature dependence of the low field magneto-conductance. Quite remarkably, weak localization effect is observed in all nanowall samples studied here. Two important parameters, namely the scattering mean free path $l_e$ and the phase coherence time $\tau_{\phi}$, are extracted by fitting the magneto-conductance data using a theory proposed by Beenakker and Houten.\cite{beenakker} The electron mobility estimated from $l_e$ is found to be comparable with those estimated previously from room temperature conductivity data for these samples\cite{bhasker1,bhasker2} confirming independently the substantial mobility enhancement in these nanowalls as compared to bulk. Like mobility, phase coherence time also shows an enhancement with the reduction of the average wall width. Interestingly, for an average wall width of 10~nm, the phase coherence time is estimated to be as high as $\approx$ 600~ps, which is much higher than those reported for GaN/AlGaN heterostructure 2DEG.\cite{Thillosen,schmult,Lehnen} Furthermore, our study of resistivity as a function of step by step etching of the walls from the top reveals that the high electron mobility region does not confine at the tip of the walls. Rather, it extends down to several hundreds of nanometer below the tip.

Nanowall network samples investigated here were grown directly on c-plane sapphire substrates using plasma assisted molecular beam epitaxial (PA-MBE) with different growth conditions. The Ga and N$_2$ fluxes were kept constant at 3.86$\times$10$^{14}$cm$^{-2}$s$^{-1}$  and 4.5 sccm, respectively. Only the substrate temperature and the duration of growth were varied as summarized in Tab.~\ref{tab1}. Further details about the growth, structural, electrical and luminescence properties of these samples can be found elsewhere.\cite{bhasker1,kesaria1}  The average width of the walls ($t_{av}$) for all of our nanowall network samples was estimated by fitting with a Gaussian the line scan profiles taken at random locations of top view SEM images recorded over different parts of the surface. The results are given in Tab.~\ref{tab1}. More details about the analysis and the distribution of wall width for these samples are provided elsewhere.\cite{bhasker2} The average electron concentration ($n$) was obtained by measuring the thermoelectric power at 300~K for these samples.\cite{bhasker1} Finally, the electron mobility [$\mu_c(300K)$] was obtained from the conductivity and electron concentration data following the model proposed in Ref. 1. The values of $n$ and $\mu_c(300K)$ thus obtained are also listed in Tab.~\ref{tab1}. Magneto transport measurements were carried out in a liquid He-cryostat equipped with a superconducting magnet in the temperature range of 1.8K to 5K. Magnetic field was varied between $\pm$ 500 Oe. Indium contact pads, which result in good ohmic contacts, were fabricated on the sample surface. Measurements were performed in four probe van-der-pauw contact geometry. 4M aqueous solution of KOH is used to etch the nanowalls for different time durations to study the depth distribution of their transport properties.


Figure~\ref{Fig1} shows the variation of the four-probe resistivity $\rho_{FP}$ as a function of the wall height of a square piece of sample F. The height of the walls is found to decrease with the duration of dipping of the sample in 4M aqueous solution of KOH. Evidently, $\rho_{FP}$ increases as the wall height decreases from 1.2~$\mu$m to 0.6~$\mu m$ in 50 minutes of dipping, implying that the top part of the walls is more conductive than the bottom part. However, it is noticeable that the enhancement is only by a few factors, meaning that the mobility is still significantly high even after the reduction of wall height by 600~nm. This suggests that the high electron mobility region does not confine only at the tip of the walls. It extends down to several hundreds of nm from the tip. The top view scanning electron microscope (SEM) images for the sample before and after 50~minutes of etching are shown in the insets of the figure. Evidently, the nanowalls forms well connected network structure, which remains intact even after the KOH treatment. However, the average tip width of the walls, which is estimated to be $\approx$ 15~nm prior to the dipping, increases to $\approx$ 50~nm after dipping, evidencing the etching of the wedge-shaped nanowalls from the top.    


Figure~\ref{Fig2}(a) compares the change in conductance $\Delta G(B)$ = $G(B)$-$G(0)$ measured at 2~K as a function of the magnetic field $B$ for different samples. Inset of the figure shows $\Delta G(B)$ plot in close-up for the samples B. In all samples, an increase of conductance with magnetic field (negative magneto-resistance) is clearly visible at low fields. This can be attributed to the weak localization effect (WL). It is noticeable that around $B$ = 0, in case of sample B and F, where $t_{av}$ is found to be only 12 and 15 nm, respectively, the rate of change of $\Delta G(B)$ with $B$ is much larger than that of sample C and D with $t_{av}$ $\approx$ 20 and 60~nm, respectively. In fact, the slope of $\Delta G(B)$ at low fields has been found to increase with the reduction of the average wall width. Figure~\ref{Fig2}(b) compares the $\Delta G(B)$ profiles recorded at 2~K for sample F before and after etching. Inset of the figure shows $\Delta G(B)$ plot in close-up for the 50~minutes etched sample. In both the cases, weak localization effect is quite evident. The slope of $\Delta G(B)$ around $B$ = 0 for the unetched sample is clearly more than that for the etched one. In fact, the slope of $\Delta G(B)$ is found to decrease as the duration of etching increases. This observation is in accordance with that of Fig.~\ref{Fig2}(a), noting that $t_{av}$ has been found to increase with etching time. 


Figure~\ref{Fig3} compares the magnetoconductance $G(B)$ profiles recorded at different temperatures for sample F (unetched).  In these temperatures, $G(B)$ increases monotonically up to a magnetic field $B_{c}$ (marked by arrows in the figure), beyond which the rate abruptly decreases. Furthermore, the change in conductance at $B_{c}$ [$\Delta G$ = $G(B_{c})$-$G(0)$] is found to decrease with the increase of temperature. Weak localization effect could not be seen in any of these samples above 3~K. Qualitatively similar results are obtained for other samples as well.

It should be mentioned that the weak antilocalization has been reported in Group III- nitride heterostructure based 2DEG systems \cite{Thillosen,schmult,Lehnen}, where it is attributed to spin-orbit coupling arising through Rashba mechanism.\cite{schmult} These results thus suggest that the spin-orbit coupling is insignificant in these samples. Note that neither weak localization nor weak antilocalization has so far been observed in bulk phase of these materials. Since the magnetic field is applied perpendicular to the sample surface in magnetoresistance measurements, all the walls are under a horizontal magnetic field at the same time as schematically shown in the inset of Fig.~\ref{Fig2}(a). Beenakker and Houten have shown that for a thin layer under a horizontal magnetic field, the quantum correction to the conductance as a function of the magnetic field can be expressed as \cite{beenakker} 

\begin{equation}
\Delta G  = -N_{ch} \frac{e^2}{2 \pi^2 \hbar} \ln \left [ \left (\frac{\tau_e}{\tau_{\phi}}+\frac{\tau_e}{\tau_{B}} \right )^{-1} + 1 \right ]
\label{eq1}
\end{equation}      

$\tau_e$ is the mean free time between two elastic collisions, $\tau_{\phi}$ the phase coherence time and $\tau_{B}$ the magnetic field dependent phase coherence time. $\tau_{B}$ = $C_1 \hbar^2$/$t_{av}^3 e^2 B^2 v_f$ in weak magnetic field regime [$l_m$ $\gg$ $\sqrt{t_{av} l_e}$], where $C_1$ = 16 and 12.1 for diffuse and specular surface scattering, respectively, $v_f$ the fermi velocity, $l_e$ = $\tau_e v_f$ the mean free path and $l_m$ = $\sqrt{\hbar/e B}$ the magnetic length. The prefactor $N_{ch}$ = 1 for a single film. However, in case of a network of walls, there are multiple channels between the contact pads, which are conducting in parallel. $N_{ch}$, which should now be the effective number of parallel channels connecting the contact pads, can thus be larger than one.  It should be noted that the above expression for $\tau_{B}$ is valid as long as $l_e$ $\gg$ $t_{av}$. The estimated values of mobility from conductance measurements [listed in Tab.~\ref{tab1}] suggest that for all nanowall network samples except sample D and the sample F after etching (Sample F-1), $l_e$ $\gg$ $t_{av}$ condition is satisfied. Sample D and F-1 belong to the dirty metal regime [$l_e$ $\ll$ $t_{av}$] with $\tau_{B}$ = $4 \hbar^2$/$t_{av}^2 e^2 B^2 D$.  The magnetoconductance profiles $G(B)$, are fitted using Eqn.~\ref{eq1} for these samples with only $N_{ch}$, $\tau_{\phi}$ and $l_e$ as fitting parameters. Here, the boundary scattering is considered to be diffusive in nature. Note that $v_f$ is obtained from the carrier concentration $n$ determined from thermoelectric power measurements [provided in Tab.~\ref{tab1}]. The fitting results for different samples are shown in Fig.~\ref{Fig3} and \ref{Fig2}. The best fit values obtained at 2~K for all the samples are listed in Tab.~\ref{tab2}. Electron mobility $\mu_{MR}(2K)$, which has been estimated from $l_e$ and $v_f$ through $\mu_{MR}(2K)=e l_e/v_f m^{*}$ with $m^{*}=0.2 m_e$ being the electron effective mass, is also listed in the table. Note that these mobility values are quite comparable with the values estimated previously from room temperature conductivity data for these samples [see Tab.~\ref{tab1}].\cite{bhasker1} Moreover, $\mu_{MR}(2K)$ increases as $t_{av}$ decreases, which is again in agreement with our previous observation.\cite{bhasker1,bhasker2} These findings not only verify the consistency of this fitting but also confirm independently the substantial increase of mobility in nanowalls as compared to bulk specially for lower wall-widths. Another noticeable point is the observation of significantly large phase coherence time ($\tau_{\phi}$) particularly for thinner walls. For instance, in sample A, $\tau_{\phi} \approx 0.6  \mu sec$, which is much higher than those reported for GaN/AlGaN heterostructure 2DEG, suggesting that the inelastic scattering rate is significantly lower in this sample as compared to heterostructure 2DEG systems.\cite{Thillosen,schmult,Lehnen}. The phase coherence length $l_{\phi}$ is also estimated following the relation $l_{\phi} = \sqrt{\tau_{\phi} D}$, where $D$ is the diffusion coefficient, for which one has Fuchs formula for a thin film with diffuse boundary scattering $D$ = $\frac{1}{3} v_f l_e \left [ 1 - \frac{3l_e}{2t_{av}} \int_o^1{s(1-s^2)(1-exp(-w/sl_e) ds} \right]$.\cite{fuchs,Sondheimer} Evidently, $l_{\phi}$ also increases as $t_{av}$ decreases. Note that for small values of $t_{av}$, $l_{\phi}$ becomes as long as 60~$\mu$m, which is higher than that is estimated in a narrow channel of GaN/AlGaN heterostructure.\cite{Lehnen}


Figure~\ref{Fig4} compares the variation of phase coherence time ($\tau_{\phi}$) as a function of temperature for different samples.  It is evident that $\tau_{\phi}$ does not vary much with temperature, suggesting the existence of a temperature independent phase breaking mechanism in this system. It has to be noted that in semiconductors, the major source of dephasing is believed to be the electron-electron scattering, which should result in a monotonic decrease of $\tau_{\phi}$ with increasing temperature. Independence of $\tau_{\phi}$ on temperature thus suggests the dominance of certain other source of dephasing than electron-electron scattering in these walls. It should be mentioned that temperature independence of $\tau_{\phi}$ has been reported in other systems such as narrow channels of GaAs/InGaAs heterostructures\cite{Hiramoto} and GaAs nanowires\cite{Hansen}. The mechanism has been ascribed to certain surface scattering process \cite{Taylor} and also to spin-spin scattering by residual magnetic impurities.\cite{Lin} It is interesting to note that $\tau_{\phi}$ increases as $t_{av}$ decreases. This could mean that the rate of inelastic scattering, which governs $\tau_{\phi}$, increases with $t_{av}$. Note that mobility is also found to increase with the decrease of $t_{av}$ for these samples. 

The observation of such a significantly high mobility and inelastic scattering time might suggest a quantum confinement of electrons in the walls. It is well known that the quantum confinement of carriers in a 2D channel can enhance the mobility by several orders of magnitude due to the decrease in scattering cross-section as a result of reduced dimensionality. The confinement in this case is likely to be 2D in nature as the high electron mobility region is found to extends down to several hundreds of nanometer below the tip of the walls (Fig.~\ref{Fig1}). It is plausible that the accumulation of certain negative charges on the nanowall surfaces can result in a 2D confinement of electrons in the wall. The surface accumulated charges on the nanowall facets might be pushing the electrons inward leading to positive depletion regions at the boundaries and a 2D quantum confinement in the central plane parallel to the height of the wall.\cite{bhasker3} Reduction of elastic as well as inelastic scattering rates with $t_{av}$ can be explained in terms of the enhancement of the 2D confinement as $t_{av}$ decreases.

In conclusion, we have studied the depth distribution of the transport properties as well as the temperature dependence of the low field magneto-conductance at low temperatures for several c-axis oriented GaN nanowall network samples grown with different average wall-widths ($t_{av}$).  Quite remarkably, weak localization effect is observed in all nanowall samples studied here. Scattering mean free path $l_e$ and the phase coherence time $\tau_{\phi}$, are extracted by fitting the magneto-conductance data using a theory proposed by Beenakker and Houten. The electron mobility estimated from $l_e$ is found to be comparable with that is estimated previously from room temperature conductivity data for these samples\cite{bhasker2}, verifying not only the consistency of this fitting but also confirming independently the substantial enhancement of mobility in these nanowalls as compared to bulk.  Interestingly, for samples with smaller wall widths, $l_{\phi}$ is estimated to be as high as 60 $\mu$m, which is much larger than those reported for GaN/AlGaN heterostructure based two dimensional electron gas (2DEG) systems. Our study , furthermore, reveals that the high electron mobility region does not confine at the tip of the walls. Rather, it extends down to several hundreds of nanometer below the tip. Both $l_e$ and $t_{\phi}$ are found to be increased with the decrease of the average wall width ($t_{av}$), suggesting a reduction of elastic as well as inelastic scattering rates with $t_{av}$ .

We acknowledge the financial support of this work by the Department of Science and Technology of the Government of India. S.\ M.\ Shivaprasad would like to thank Prof. C.\ N.\ R.\ Rao for his support and guidance.

\newpage
\phantom{000}
\begin{table}[t!]
\centering
\caption{The average wall width $t_{av}$, growth temperature $T_{s}$, growth time $t_{gr}$, carrier concentration $n$ and   electron mobility $\mu$ are listed for these samples.}
\begin{tabular}{c|c|c|c|c|c} 
\hline
\hline

Sample & $t_{av}$ & $T_{s}$  & $t_{gr}$               & $n$                                & $\mu$                    \\ 
       & (nm)     & (\celsius)  & (hours)             & $\times$ 10$^{19}$ (cm$^{-3}$)     &(cm$^2$/V s)              \\ \hline	
A     & 10 $\pm 5.9$       &   630       & 4          & 1.25                               & 2.05$\times$ 10$^{4}$    \\ 
B     & 12 $\pm 3.3$       &   650       & 2          & 1.19                               & 1.008$\times$ 10$^{4}$   \\
F     & 15 $\pm 3.8$       &   640       & 4          & 1.23                               & 1.1$\times$ 10$^{4}$      \\  
C     & 18 $\pm 5.3$       &   530       & 2          & 1.17                               & 7.95$\times$ 10$^{2}$     \\ 
F-1$^{a}$ & 40 $\pm 10$    &  640        & 4          & 1.30                               & 5.49$\times$ 10$^{2}$        \\ 
                                                                                                              
D     & 60 $\pm 10$        &   780       & 2          & 1.32                               & 28            \\

\hline 
\hline
\end{tabular}
\label{tab1}
\end{table} 
$^{a}$ ~Sample F after 50 minutes of etching

\newpage
\phantom{000}
\begin{table}[t!]
\centering
\caption{The average wall width $t_{av}$, effective number of channels connecting between two Indium metal pads $N_{ch}$,elastic mean free path $\l_e$, elastic mean time $\tau_{e}$, fermi velocity $v_f$, electron mobility estimated from $\l_e$ and $v_f$ using $\mu_{MR}(2K)=e l_e/v_f m^{*}$, phase coherence time $\tau_{\phi}$ and phase coherence length $l_{\phi}$ are given.}
\begin{tabular}{c|c|c|c|c|c|c|c|c} 
\hline
\hline
Sample & $t_{av}$ & $N_{ch}$   & $l_e$ & $\tau_{e}$    & $v_{f}$     & $\mu_{MR}$                   & $\tau_{\phi}$        & $l_{\phi}$  \\ 
     & (nm)  & (cm$^{-1}$)     & (nm)   & $\times$ 10$^{-12}$(s) &$\times$10$^5$(M/s)& (cm$^2$/V s) & $\times$ 10$^{-10}$ (s)   & ($\mu$m)  \\ \hline
 A$^{b}$ & 10$\pm 5.9$   & 1400   &  2080 $\pm 208$   & 4.96  & 4.19  &43590 & 6080$\pm 608$                               & 60.6 $\pm 6.1$  \\ 
 B   & 12$\pm 3.3$   & 2750   &  1130 $\pm 113$   & 2.8   & 4.05  &24524 & 945  $\pm 95$                                   & 23.8 $\pm 2.4$ \\  
 F   & 15$\pm 3.8$   & 3000   &  723  $\pm 72.3$  & 1.78  & 4.05  &15646 & 714  $\pm 71$                                   & 21.6 $\pm 2.2$ \\   
 C   & 18$\pm 5.3$   & 4000   &  28   $\pm 2.8$   & 0.07  & 4.045 &610   & 1.96 $\pm 0.6$                                  & 0.65$\pm 0.07$ \\ 
 F-1 & 40$\pm 10$    & 3000   &  25   $\pm 2.5$   & 0.06  & 4.08  &549   & 1.25 $\pm 0.13$                                 & 0.57 $\pm 0.06$\\ 
 D   & 60$\pm 10$    & 4410   &  1    $\pm 0.1$   & 0.003 & 4.053 &24    & 6.2  $\pm 0.62$                                 & 0.30 $\pm 0.03$ \\ 
\hline 
\hline
\end{tabular}
\label{tab2}
\end{table}  
$^{b}$ ~ measured at 2.4K

\newpage
\begin{figure}[ht!]
\caption{ Variation of the four-probe resistivity $\rho_{FP}$ of a square piece of sample F as a function of the wall height. Note that the height of the walls is found to decrease with the duration of dipping in 4M aqueous solution of KOH.  Inset shows the top view SEM images of samples F before and after 50 minutes of etching.}
\label{Fig1}
\end{figure}

\begin{figure}[ht!]
\caption{(a) Compares the change in conductance $\Delta G(B)$ = $G(B)$-$G(0)$ measured at 2~K as a function of the magnetic field $B$ for different samples. Left inset of the figure shows $\Delta G(B)$ plot in close-up for the samples B. Position of $B_{c}$ is indicated by the arrows. Right inset schematically shows the orientation of magnetic field with respect to the orientation of the walls (b)  Compares the $\Delta G(B)$ profiles recorded at 2~K for sample F before and after etching. Inset of the figure shows $\Delta G(B)$ plot in close-up for the 50~minutes etched sample. Solid line show the fitting to the experimental data using equation no. ~1.}
\label{Fig2}
\end{figure}

\begin{figure}[ht!]
\caption{Compares the magnetoconductance $G(B)$ profiles recorded at different temperatures for sample F (unetched). Solid line show the fitting to the experimental data using equation no. ~1.}
\label{Fig3}
\end{figure}

\begin{figure}[ht!]
\caption{Compares the variation of phase coherence time ($\tau_{\phi}$) as a function of temperature for different samples.  }
\label{Fig4}
\end{figure}

\newpage
\phantom{000}
\vfill
\centerline{\includegraphics*[width=12cm]{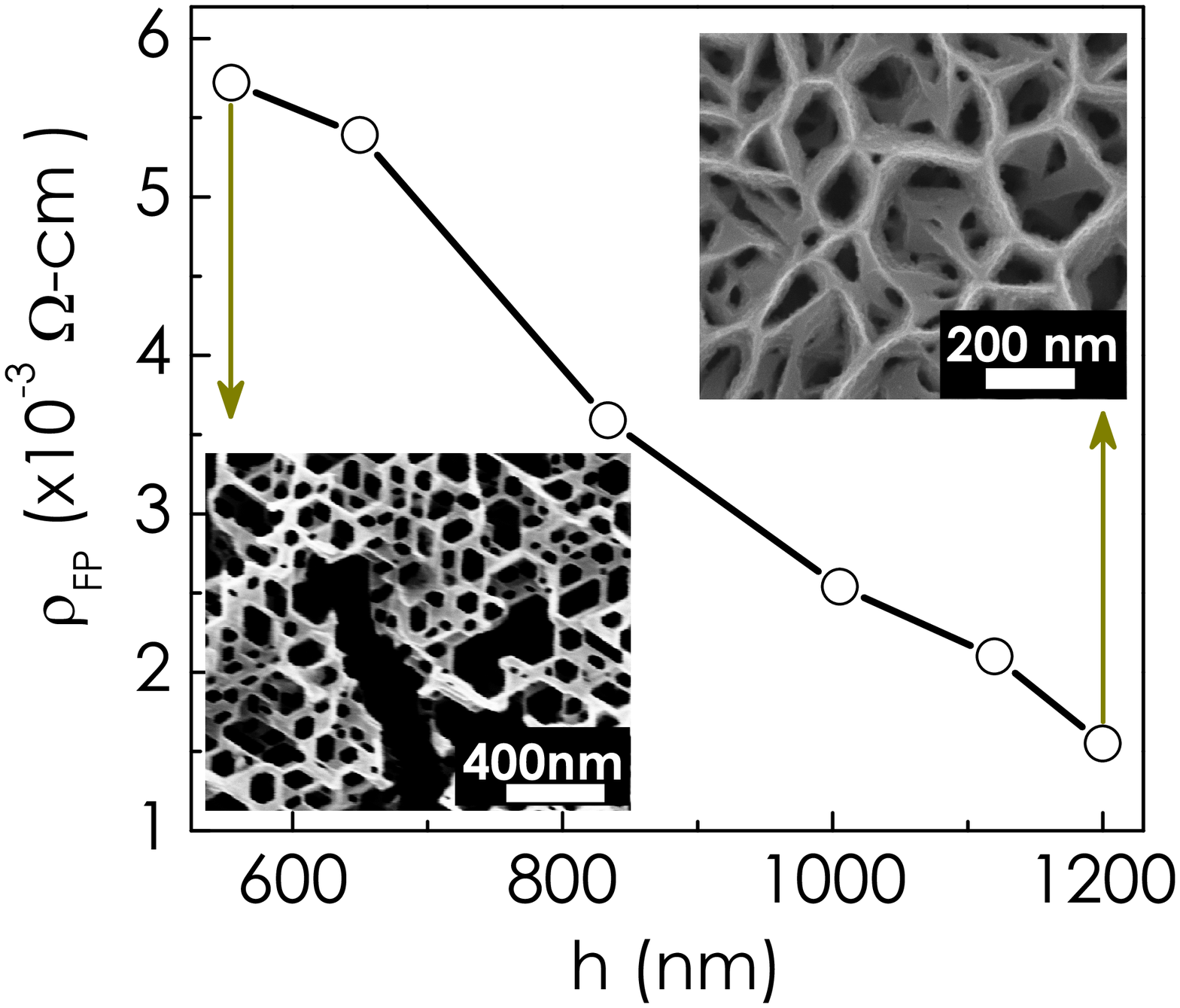}}
\vfill
\textsf{\large Fig.~1 of Bhasker \textsl{et al.}}

\newpage
\phantom{000}
\vfill
\centerline{\includegraphics*[width=12cm]{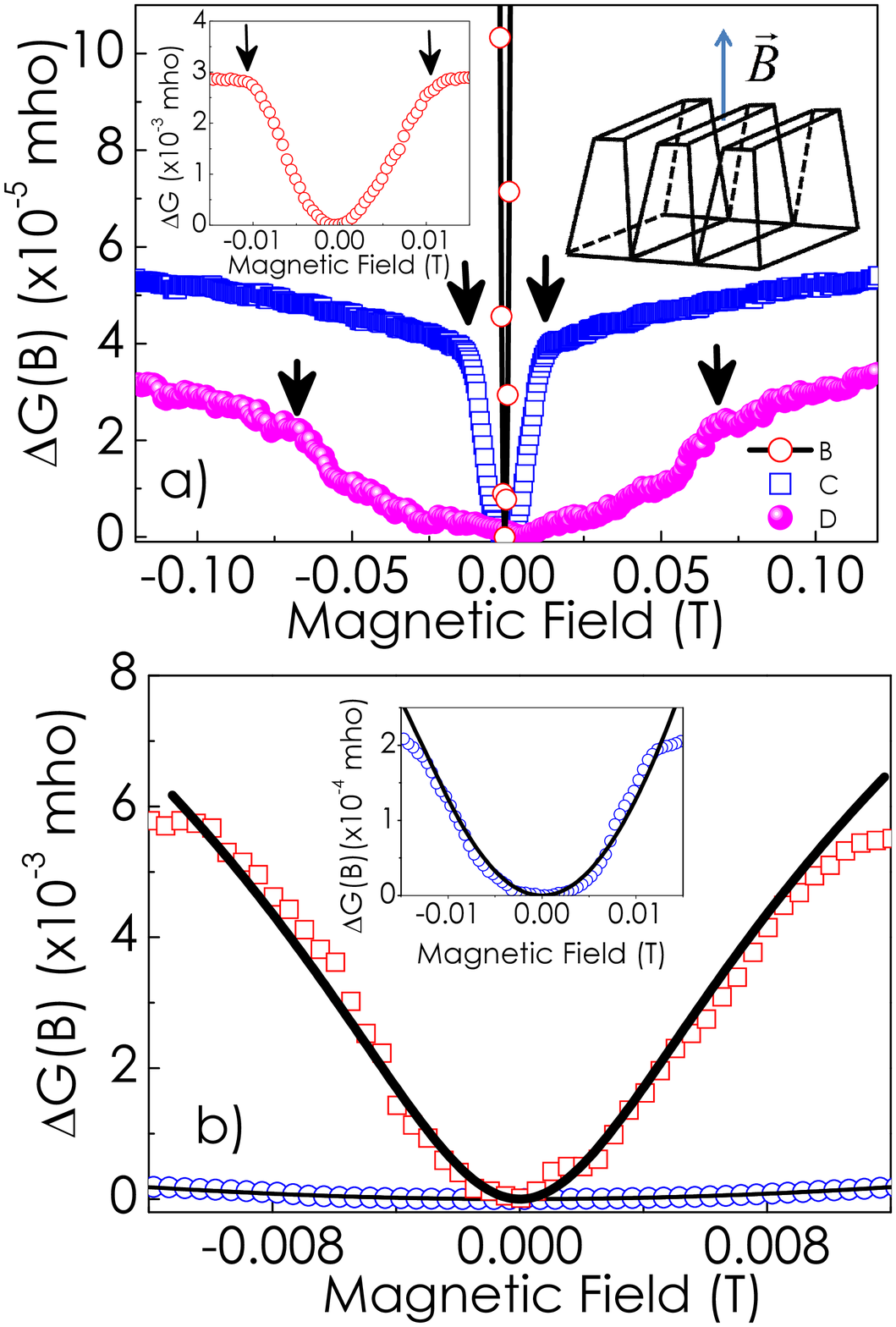}}
\vfill
\textsf{\large Fig.~2 of Bhasker \textsl{et al.}}

\newpage
\phantom{000}
\vfill
\centerline{\includegraphics*[width=12cm]{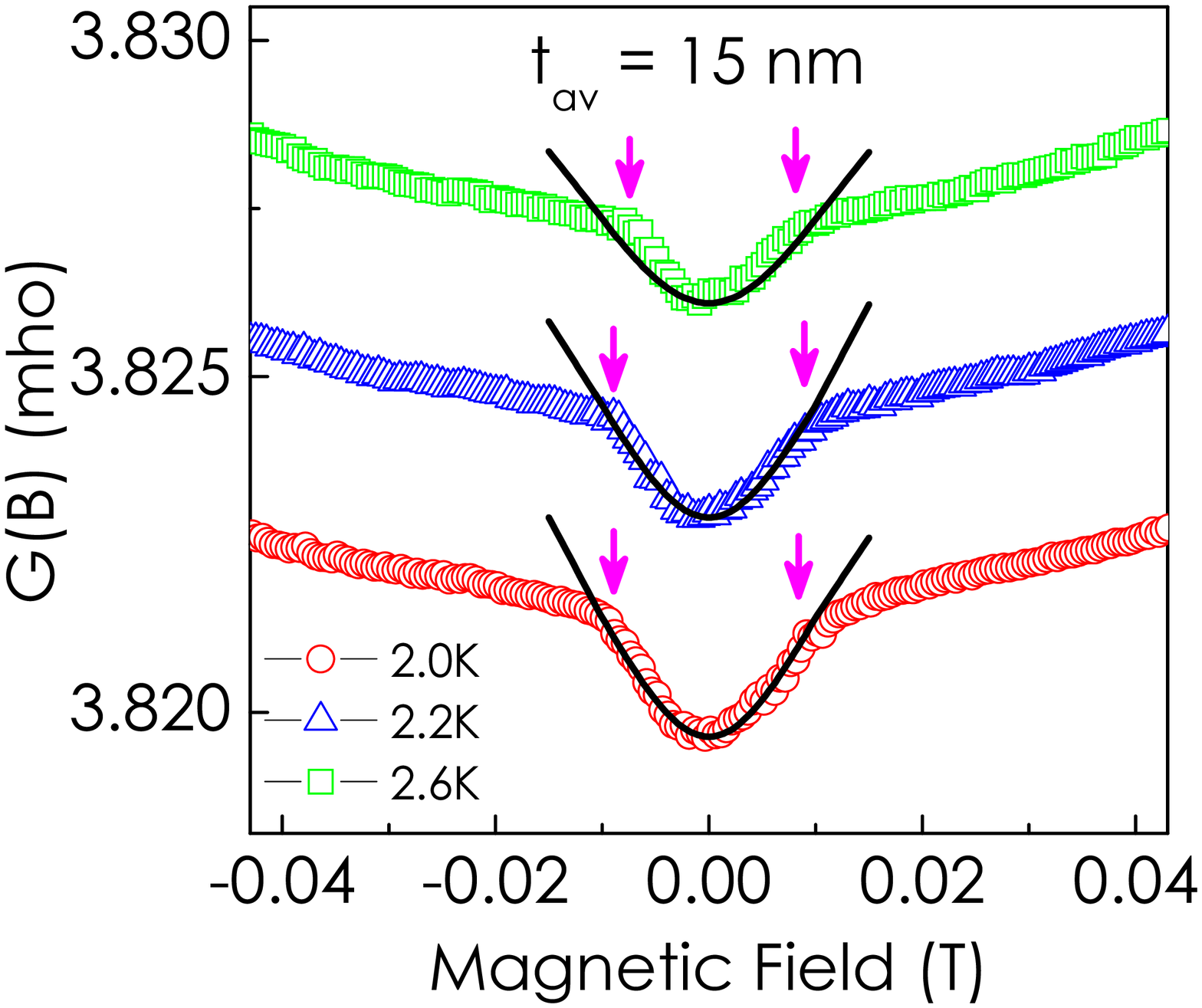}}
\vfill
\textsf{\large Fig.~3 of Bhasker \textsl{et al.}}

\newpage
\phantom{000}
\vfill
\centerline{\includegraphics*[width=12cm]{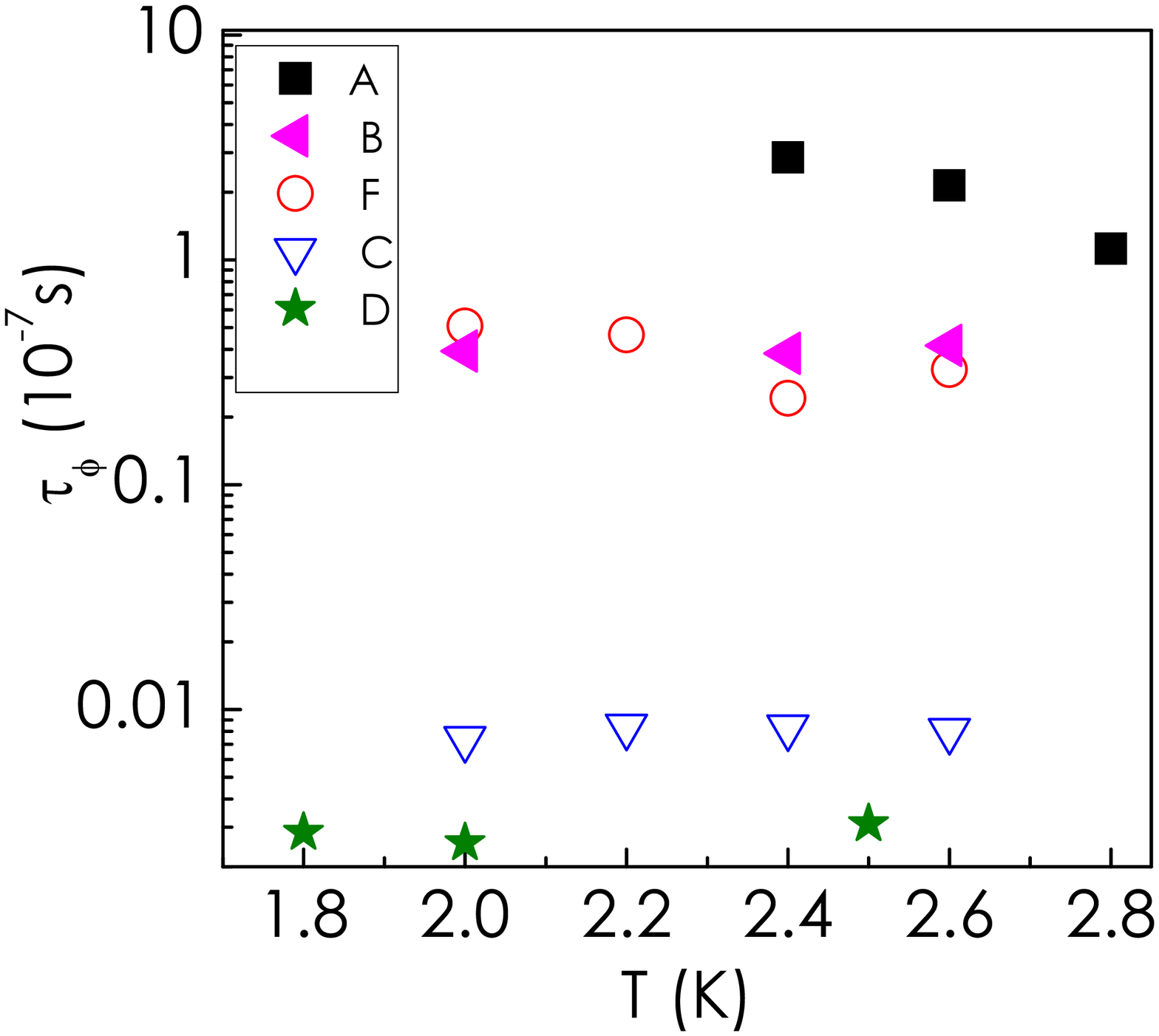}}
\vfill
\textsf{\large Fig.~4 of Bhasker \textsl{et al.}}
\end{document}